\documentclass{article}
\usepackage[english]{babel}
\usepackage{amsmath}
\usepackage{amssymb}
\usepackage{graphicx}
\usepackage[colorlinks=false, hidelinks]{hyperref}
\usepackage{multirow}
\usepackage{csquotes}

\usepackage[
    backend=biber,
    style=apa,
    sorting=nyt,
    giveninits=true,
    uniquename=init,
    natbib=true
]{biblatex}
\DeclareFieldFormat{bibhyperrefnonest}{%
  \DeclareFieldFormat{bibhyperref}{##1}%
  \bibhyperref{#1}}

\DeclareCiteCommand{\parencite}[\mkbibparens]
  {\usebibmacro{cite:init}%
   \usebibmacro{prenote}}
  {\usebibmacro{citeindex}%
   \printtext[bibhyperrefnonest]{\usebibmacro{cite}}}
  {}
  {\usebibmacro{postnote}%
   \usebibmacro{cite:post}}

\title{Smart Sampling Strategies for Wireless Industrial Data Acquisition \\[1ex] \large Minimizing Aliasing While Preserving Trends for Machine Learning Models}
\author{Soto Marcos \\[1ex] \normalsize\textit{Universidad Loyola Andalucía}}
\date{}

\begin{document}

\maketitle

\begin{abstract}
    In industrial environments, data acquisition accuracy is crucial for process control and optimization. Wireless telemetry has proven to be a valuable tool for improving efficiency in well-testing operations, enabling bidirectional communication and real-time control of downhole tools \parencite{acar_optimizing_2014}.
    There is a growing adoption of wireless systems due to their advantages. However, the high sampling frequencies required present significant challenges in telemetry, including data storage, data transmission, computational resource consumption, and battery wear of wireless devices \parencite{julius_wireless_nodate}.
    This study explores how applying mathematical techniques for optimizing a real data acquisition system can detect aliasing effects, and systematic errors, and improve sampling rates without compromising the accuracy of physical measurements.
    An 80\% reduction in the sampling frequency was achieved without compromising measurement quality. These findings provide a basis for optimizing the implementation of wireless data acquisition systems in industrial environments.    
\end{abstract}

\section{Introduction}

As a fundamental task in the oil exploration industry, well testing is a process in which the production of a well is measured over a specified period, either at the wellhead using portable equipment or at a production facility. Its primary purpose is to evaluate key reservoir parameters such as permeability, hydraulic connectivity, and average reservoir pressure.
In exploratory wells, it is used to collect fluid samples, measure initial pressure, and estimate the reservoir's minimum volume. In production wells, it allows verifying permeability, identifying fluid behaviors, and evaluating heterogeneities. Regarding the produced phases or fluids from this operation, liquids such as water and oil are stored in atmospheric tanks, while gas is usually flared or delivered to a sales point.
This process is crucial to ensure proper reservoir characterization, optimize production, and minimize errors in operational and strategic decision-making. Accurate measurements ensure that reservoir models reflect real conditions more precisely, maximizing project profitability and minimizing risks \parencite{spe_welltest_2019}.
This study focuses on analyzing the relationship between sampling frequency \( F_s \) and the quality of a measured signal, evaluated through the relative error between the original signal and the reconstructed signal. In data acquisition systems, choosing an appropriate sampling frequency is crucial to ensure measurement accuracy while minimizing storage and transmission costs. However, lower sampling frequencies can introduce aliasing errors, affecting the quality of signal reconstruction.
Aliasing is a phenomenon that occurs when the sampling frequency is insufficient to accurately capture the variability of a signal, causing important patterns to be distorted or even lost. In industrial systems, this can be critical as it may prevent early detection of anomalies in the data, affecting the ability of machine learning (ML) models to prevent failures or detect key trends. Therefore, optimizing the sampling rate is not just about reducing data acquisition but about ensuring that downsampled signals retain their essential characteristics for predictive analytics.
In this context, the specific case of gas measurement is investigated, excluding liquid measurement as it can be contrasted with accumulated tank volume. In contrast, gas must be measured in real time since, once vented or burned, it cannot be subsequently contrasted. Therefore, the goal is to investigate whether it is possible to reduce the sampling frequency using compensation techniques to maintain the error within acceptable limits.
Some previous studies have explored techniques for energy-efficient data acquisition,\parencite{law_energy-efficient_2009} adapting industrial sampling frequency, but not in their specific application to reduce aliasing in signals allowing their behavior patterns to be maintained in gas measurements in well-testing environments. This would allow optimizing the resources associated with data storage and transmission without compromising the quality of the measurements.

\section{Objective}
This study aims to determine the minimum sampling frequency that optimizes data storage and transmission in wireless acquisition systems while ensuring that the signal quality is maintained. Specifically, the goal is to avoid aliasing errors that could affect the ability of ML models to identify trends and detect anomalies in industrial environments. By striking a balance between efficient data acquisition and signal preservation, we ensure that the reduced sampling frequency does not compromise the effectiveness of predictive models used in industrial monitoring.

To achieve the study's objective, the aim is to determine the optimal sampling frequency \( F_s \) that minimizes the costs associated with the data acquisition system while maintaining a relative error \( E_{\text{relative}} \) below a target value \( E_{\text{objective}} \). To this end, the following objectives are proposed:
\begin{itemize}
    \item Evaluate the impact of \( F_s \) on the relative error of the uncompensated signal.
    \item Implement compensation techniques to reduce relative error at lower sampling frequencies.
    \item Define a cost function that combines the impact of relative error with storage and transmission costs.
    \item Solve an optimization problem to identify the minimum sampling frequency that meets the established quality criteria.
\end{itemize}

It is crucial to ensure measurement quality to maintain a functional dynamic system with sufficient representation of measured magnitudes. For example, if anomaly detection models rely on measured trends, discretization errors that significantly alter the signal representation can negatively impact such dependent models.

\section{Methodology}
\subsection{Mathematical Framework}
Relative error: It is defined as the main metric to evaluate the quality of the signal reconstruction:
\[
E_{\text{relative}} = \frac{\| S_{\text{original}} - S_{\text{rebuilt}} \|}{\| S_{\text{original}}\|}
\]
Where \( S_{\text{original}} \) is the original signal and \( S_{\text{rebuilt}} \) is the reconstructed signal after subsampling and applying compensation techniques.

\subsection{Cost function}
In wireless systems, the transmission cost is directly related to the energy consumption of the sensor, while the storage cost depends on the amount of data generated, proportional to the sampling frequency \( F_s \). Therefore, the cost function is posed as:
\[
C(f_s) = k_a f_s + k_t E\text{trans} (f_s) + \lambda E_{\text{relative}}
\]
Where:
\begin{itemize}
    \item \( k_a \): Constant representing the cost proportional to data storage.
    \item \( k_t \): Constant that represents the cost associated with energy consumption per transmission.
    \item \( E_{\text{trans}}(f_s) \): Energy consumption associated with data transmission at a sampling frequency \( f_s \).
    \item \( \lambda \): Weight that reflects the relative importance of the error in the total cost function.
\end{itemize}

For wireless sensors, \( E_{\text{trans}}(f_s) \) can be approximated as a proportional to the number of transmissions:
\[ 
E_{\text{trans}}(f_s) = f_s \times E_{\text{unit}} 
\]

Where \( n_{\text{trans}} \) is the number of transmissions in period \( T \) at a sampling frequency \( f_s \), and \( E_{\text{unit}} \) is the energy consumption per transmission.

Quality constraint: To ensure that the signal quality is acceptable, the following is set:
\[
E_{\text{relative}} \leq E_{\text{target}}
\]

\subsection{Relationship between frequency and relative error}
According to the Nyquist criterion, the sampling frequency should be in the range:
\[
f_s \in \left[f_{\text{min}}, 2 f_{\text{max}}\right]
\]
where \( f_{\text{min}} \) is the minimum acceptable sampling frequency according to the target error \( E_{\text{target}} \).

The aliasing error is evaluated as the accumulated difference between the original and reconstructed signals in the frequency domain:
\[
E_{\text{aliasing}} = \int_{f_s}^{2 f_{\text{max}}} \left| S_{\text{original}}(f) - S_{\text{rebuilt}}(f) \right| \, df
\]

This allows us to evaluate the magnitude of the error when reducing \( f_s \), considering its impact on the quality of the reconstructed signal.\parencite{aljameel_anomaly_2022}

\subsection{Experimental Setup}
The experimental process consists of the following steps:

\subsection{Data Collection}
Data collection: Real-time data is collected from a wireless sensor network in an industrial environment, focusing on gas measurements in well-testing operations.
Several datasets collected from different time periods and different oil wells were used, obtained from well testing services at oil and gas companies. The data includes gas, oil and water flow measurements. The data was originally sampled at a frequency of one data per second. This high frequency allows accurate capture of signal variability, but presents challenges in terms of storage and power consumption of wireless instruments.

\subsubsection{Description of the experiments}

Purpose: Subsampling with different frequencies will be performed on real field signals. The experiments seek to validate how the relative error varies with the sampling frequency and how compensation techniques allow maintaining signal quality when reducing the frequency.
Signals used: 
For the experiment, two signals were taken from two different oil wells, in a gas flow measurement during a full day of a real field operation. These signals correspond to the gas line measurement that goes from a well testing separator to the vent line.
Sampling frequencies: The signals were subsampled at frequencies of 1, 5, 10, 15, and 20 seconds. The original signal was sampled at 1 Hz.

\begin{figure}[h!]
    \centering
    \includegraphics[width=0.8\textwidth]{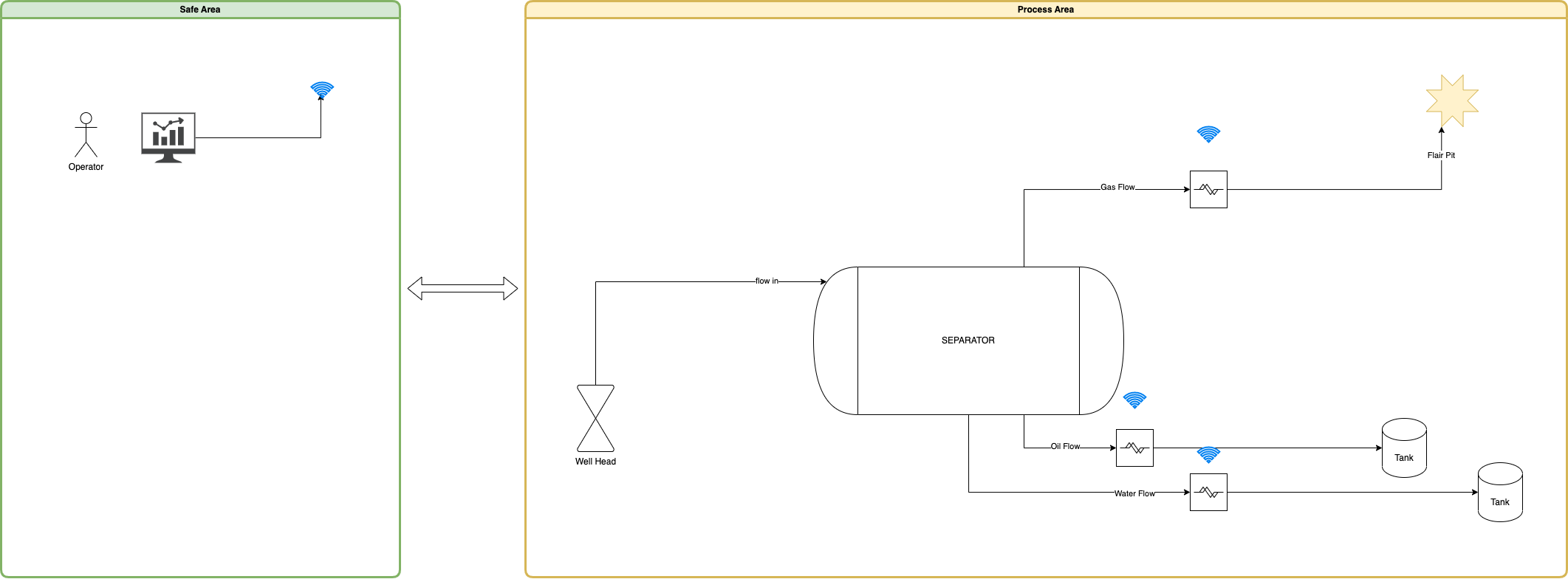}
    \caption{Data Acquisition Process}
    \label{fig:original_signal}
\end{figure}

A wireless sensor was installed on the flow line, the process area is located at a safe distance from the control room, the data is sent wirelessly.
The signals of gas volumetric flow rate as a function of time are shown below:

\begin{figure}[h!]
    \centering
    \includegraphics[width=0.8\textwidth]{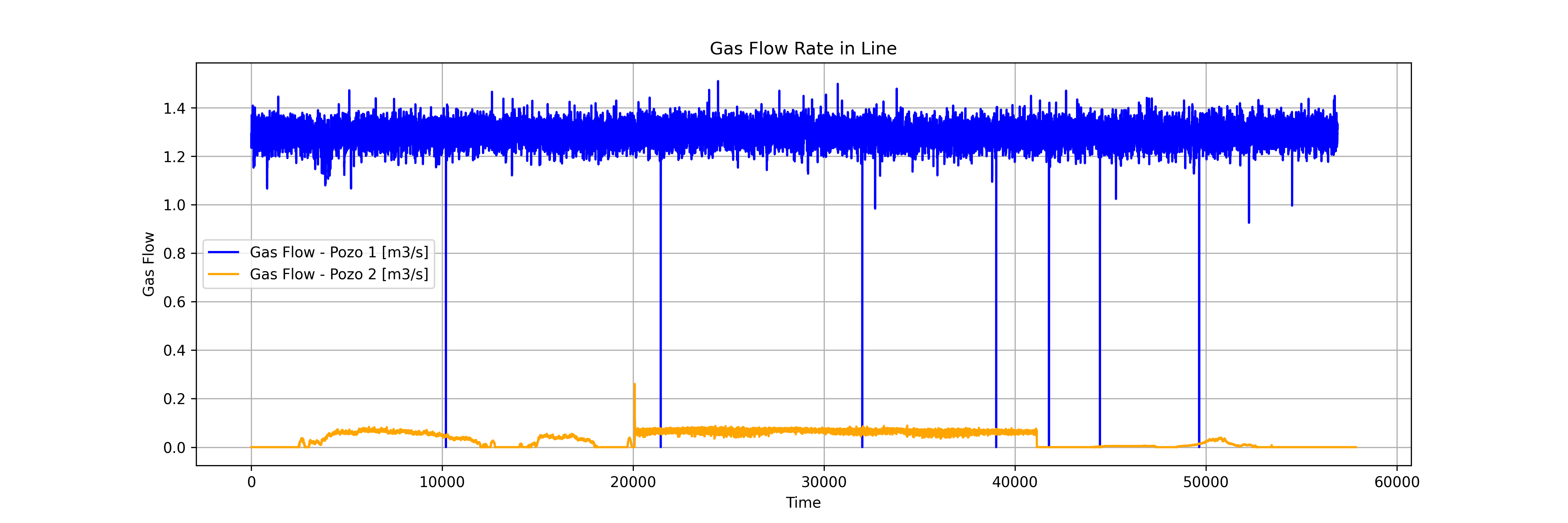}
    \caption{Gas Flow Signals}
    \label{fig:gas_flow_signals1}
\end{figure}

It can be observed that the signal corresponding to Well number 1 has a fairly repetitive behavior pattern, while the signal corresponding to Well 2 does not. This may be due to characteristics of the operation, such as moments of changes in the well's production, closing/opening of the well, communication problems and/or possible unexpected events.

\begin{figure}[h!]
    \centering
    \includegraphics[width=0.8\textwidth]{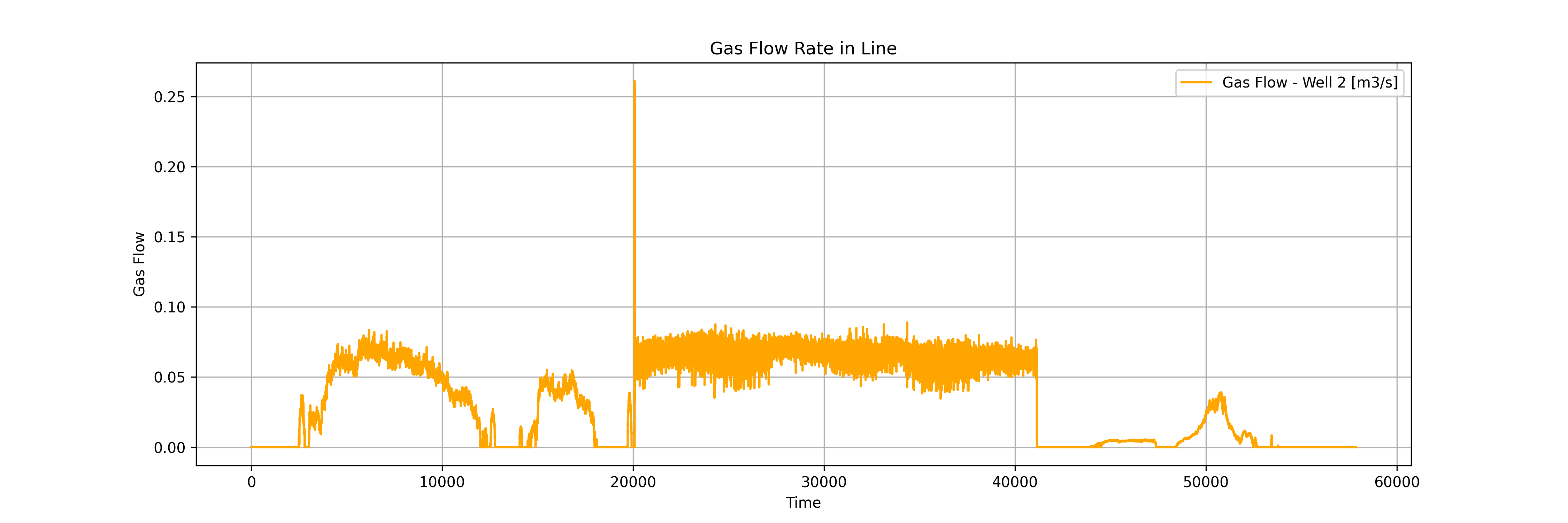}
    \caption{Gas Flow Signal Well Nro 2}
    \label{fig:gas_flow_signals2}
\end{figure}

As part of the measurement task, it is important to detect these events in time to prevent any potential danger in the installation. Just as an example, a sharp drop in the gas trend can mean a blockage in the vent outlet that can cause an explosion or rupture of the pipe with high risk potential.

\subsection{Analysis Methods}
To identify the magnitude of the error, among the signals that were obtained using different frequencies, we can take different techniques:

\begin{itemize}
    \item Compare the entire shape of the signal.
    \item Compare only the difference in means.
\end{itemize}

The first technique is more sensitive to changes in the signal, while the second is more robust to noise. In this case, the first technique was used to evaluate the error magnitude.

\subsubsection{Signal Comparison Techniques}

\paragraph{Compare the entire shape of the signal}
\noindent
\vspace{0.25cm}

For this purpose, it is common to calculate the norm (magnitude) of a vector (Euclidean or L2 norm). This means that, given a list (or array) of values, for example, \([x_1, x_2, x_3, \dots]\), it returns:
\[
\sqrt{x_1^2 + x_2^2 + x_3^2 + \dots}
\]

In other words, it is the Euclidean distance of the vector from the origin.

This metric, the Euclidean or L2 norm, evaluates the point-to-point difference between the original signal and the reconstructed signal. The more “similar” they are at all points, the lower this value will be.

\paragraph{Interpretation:}
\noindent
\vspace{0.25cm}

It takes into account not only the average amplitude, but also the complete shape of the signal (peaks, valleys, phase, etc.).
Ideal for measuring the overall similarity of the waveform (for example, when the shape of the signal is crucial: ECG, vibrations, etc.).

drawbacks:
\noindent
\vspace{0.25cm}

If only the average magnitude (average amplitude) is of interest, the error may be “over-measured”, since a small sustained variation may not be so relevant for the application.

\paragraph{Compare only the difference in means:}
\noindent
\vspace{0.25cm}

When you are only interested in the average amplitude, you can compare the mean of the original signal and compare it against the mean of the downsampled signal.
This measure how different the average amplitude of the reconstructed signal is from the original.

\paragraph{Interpretation:}
\noindent
\vspace{0.25cm}

If the signal mean is the only target (e.g., an offset or power average), this metric may be sufficient and easier to interpret.
It does not reflect shape differences or point variations, only the overall shift in average magnitude.

drawbacks:
\noindent
\vspace{0.25cm}

It can mask large point differences across the signal. For example, the mean may match even if there are very different peaks in shape or phase.

\subsubsection{Compensation Methods}
\paragraph{Application of compensation techniques:}
\noindent
\vspace{0.25cm}

For each of the signals, well 1 and 2, different sampling frequencies (\(f_s\)) are applied, that is, the signal is subsampled using an appropriate factor, and the L2 error is calculated, and with respect to the mean, even taking different values for the variable \(e\) (allowable error).

\paragraph{Two sets of experiments were performed:}
\noindent

\begin{itemize}
    \item Experiment 1: Without compensation: Using the original signal.
    \item Experiment 2: With compensation: Using the filtered and reconstructed signal to mitigate the aliasing error. \parencite{martinez_nuevo_nonuniform_2021}
\end{itemize}

\section{Results and Discussion}
\subsection{Uncompensated Signal Analysis}
\subsubsection{Experiment 1: Without compensation}

\begin{figure}[h!]
    \centering
    \includegraphics[width=0.8\textwidth]{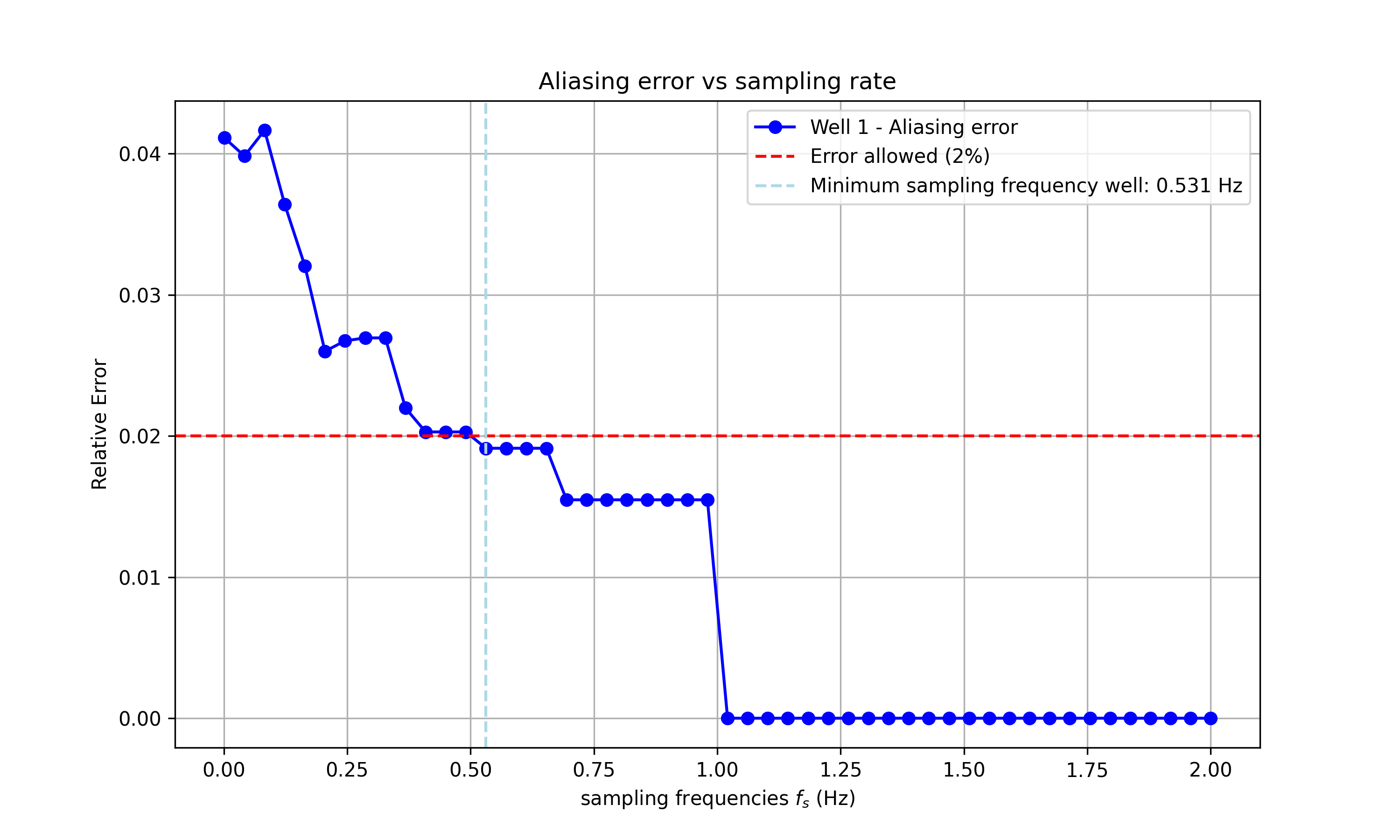}
    \caption{Well 1, relationship between L2 error and sampling rate (\(f_s\)) for the original signal, showing a significant increase in error at low frequencies due to aliasing.}
    \label{fig:nc_well1}
\end{figure}

\begin{figure}[h!]
    \centering
    \includegraphics[width=0.8\textwidth]{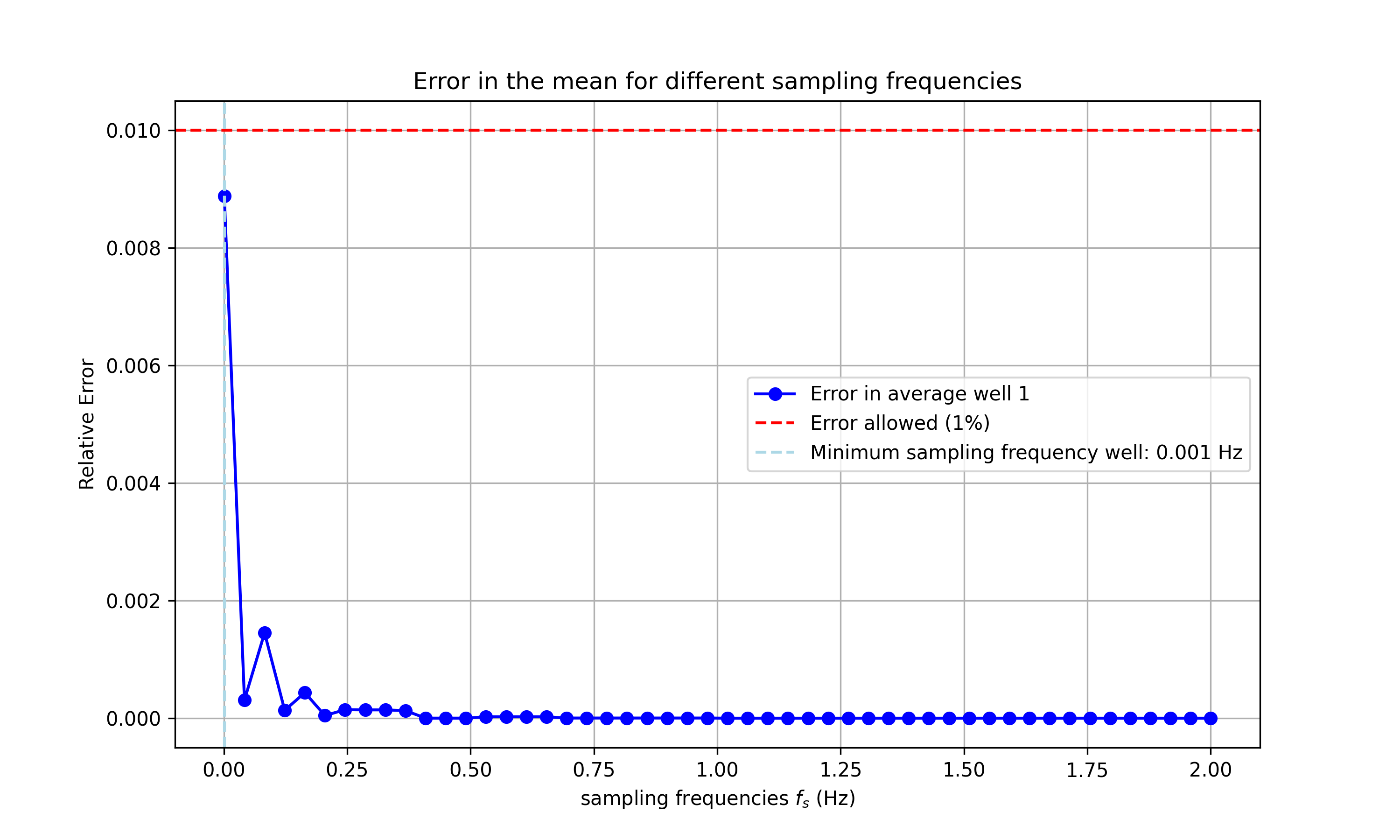}
    \caption{Well 1, relationship between the error relative to the mean and the sampling frequency (\(f_s\)) for the original signal.}
    \label{fig:nc_well1_mean}
\end{figure}

\begin{figure}[h!]
    \centering
    \includegraphics[width=0.8\textwidth]{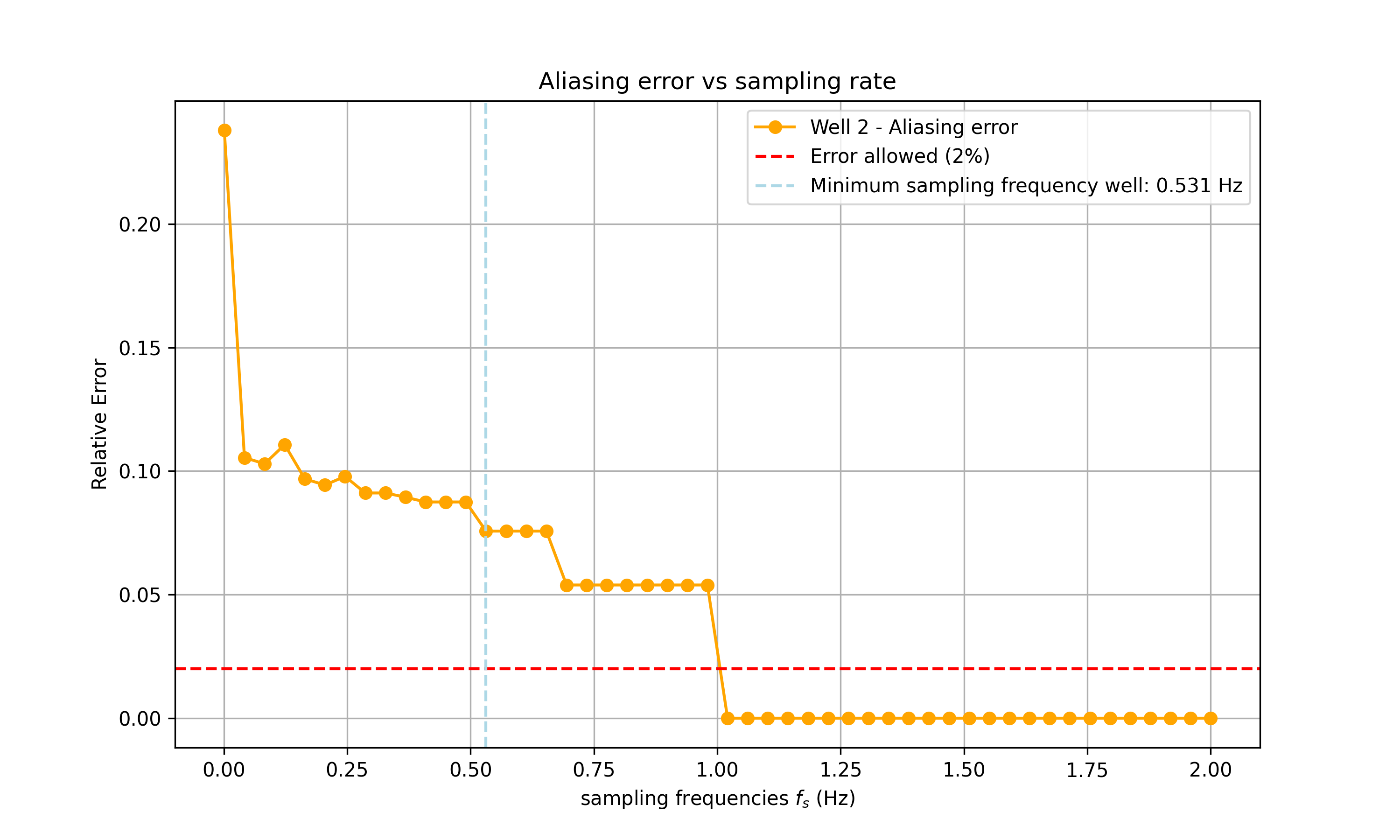}
    \caption{Well 2, relationship between L2 error and sampling rate (\(f_s\)) for the original signal, showing a significant increase in error at low frequencies due to aliasing.}
    \label{fig:nc_well2}
\end{figure}

\begin{figure}[h!]
    \centering
    \includegraphics[width=0.8\textwidth]{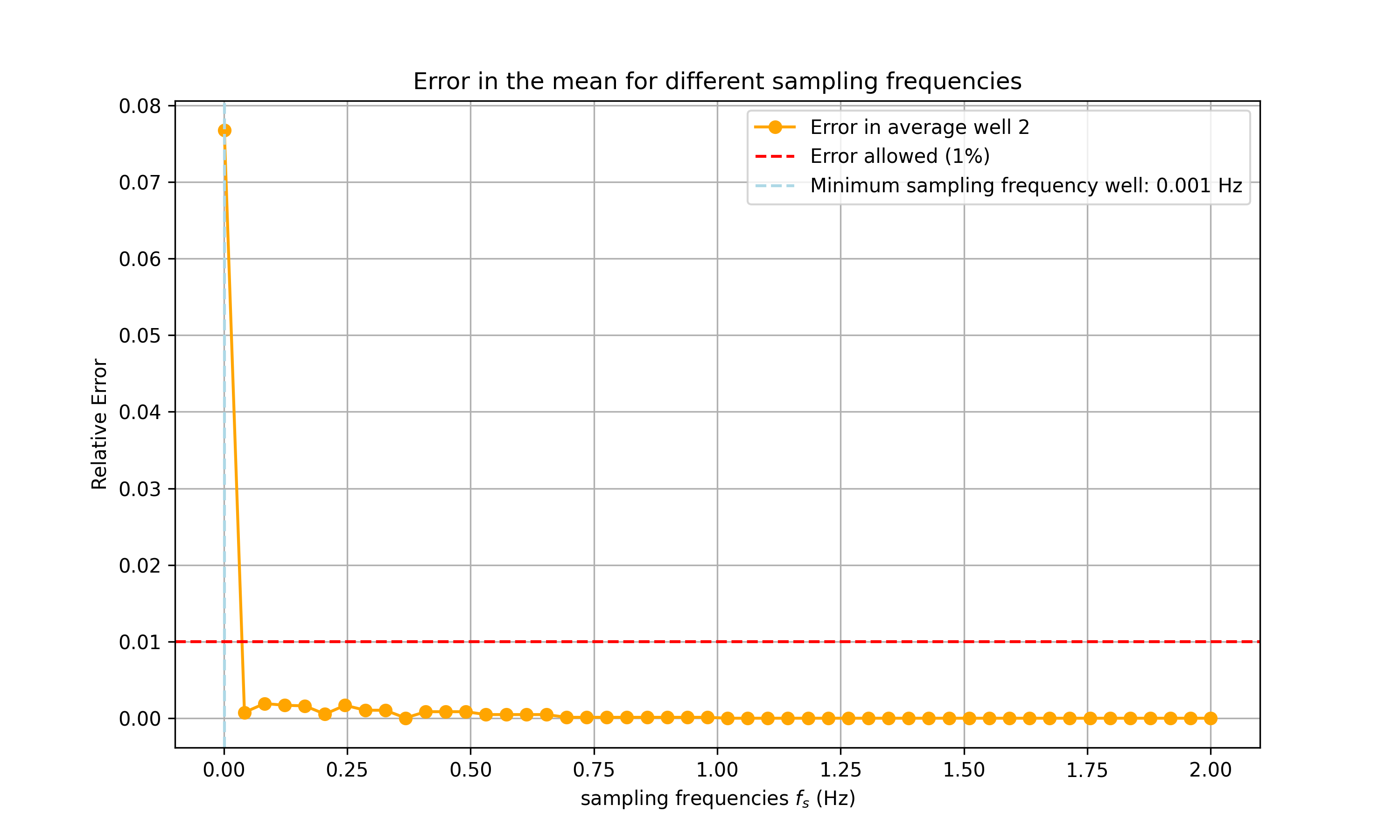}
    \caption{Well 2, relationship between the error relative to the mean and the sampling frequency (\(f_s\)) for the original signal.}
    \label{fig:nc_well2_mean}
\end{figure}

\subsection{Compensated Signal Analysis}
\subsubsection{Experiment 2: With compensation}

Compensation techniques were applied to reconstruct the signal using cubic interpolation and outlier filtering.

\begin{figure}[h!]
    \centering
    \includegraphics[width=0.8\textwidth]{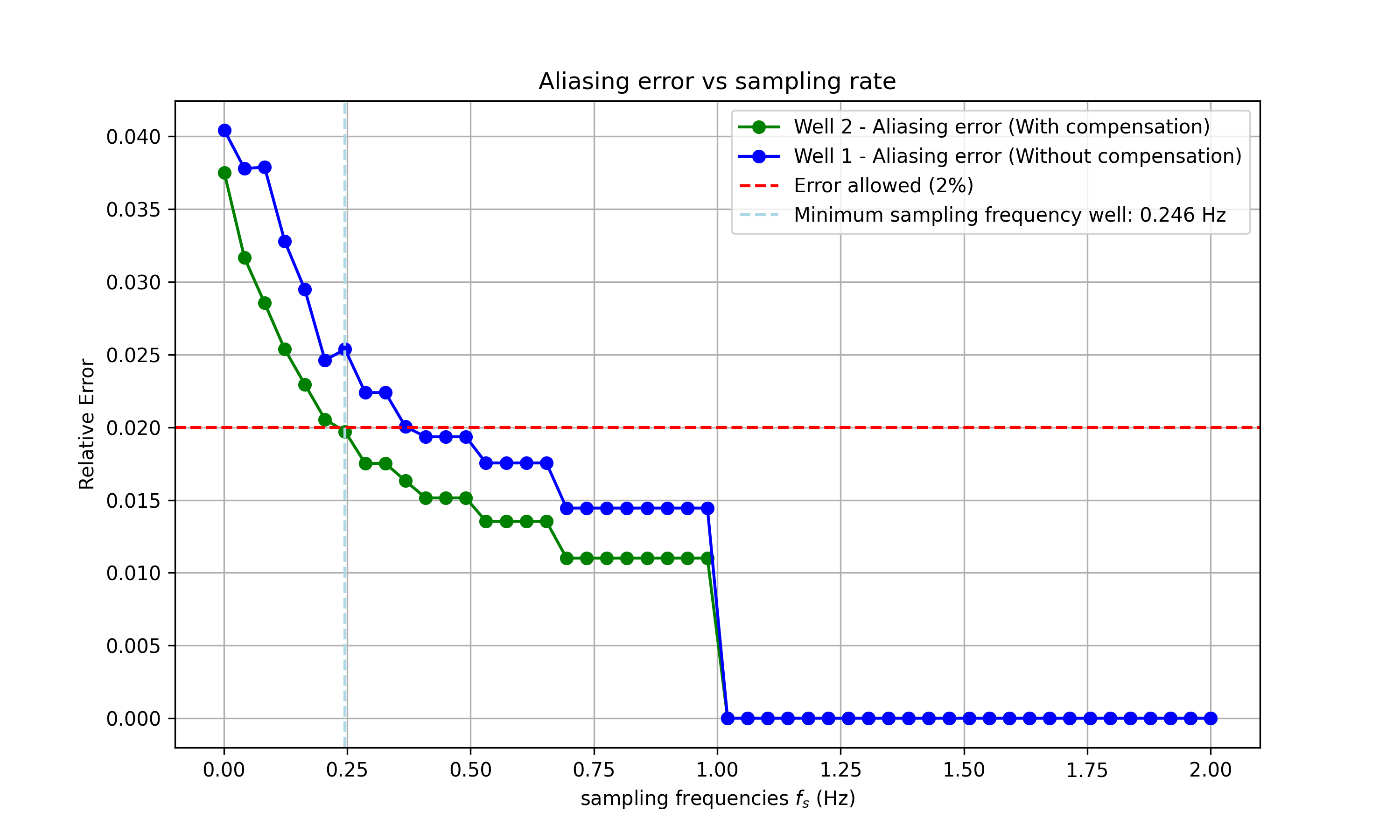}
    \caption{Well 1, relationship between L2 error and sampling rate (\(f_s\)) with compensation techniques (filtering + cubic reconstruction), showing that the compensated signal can be used at lower sampling rates than the original one.}
    \label{fig:c_well1}
\end{figure}

\begin{figure}[h!]
    \centering
    \includegraphics[width=0.8\textwidth]{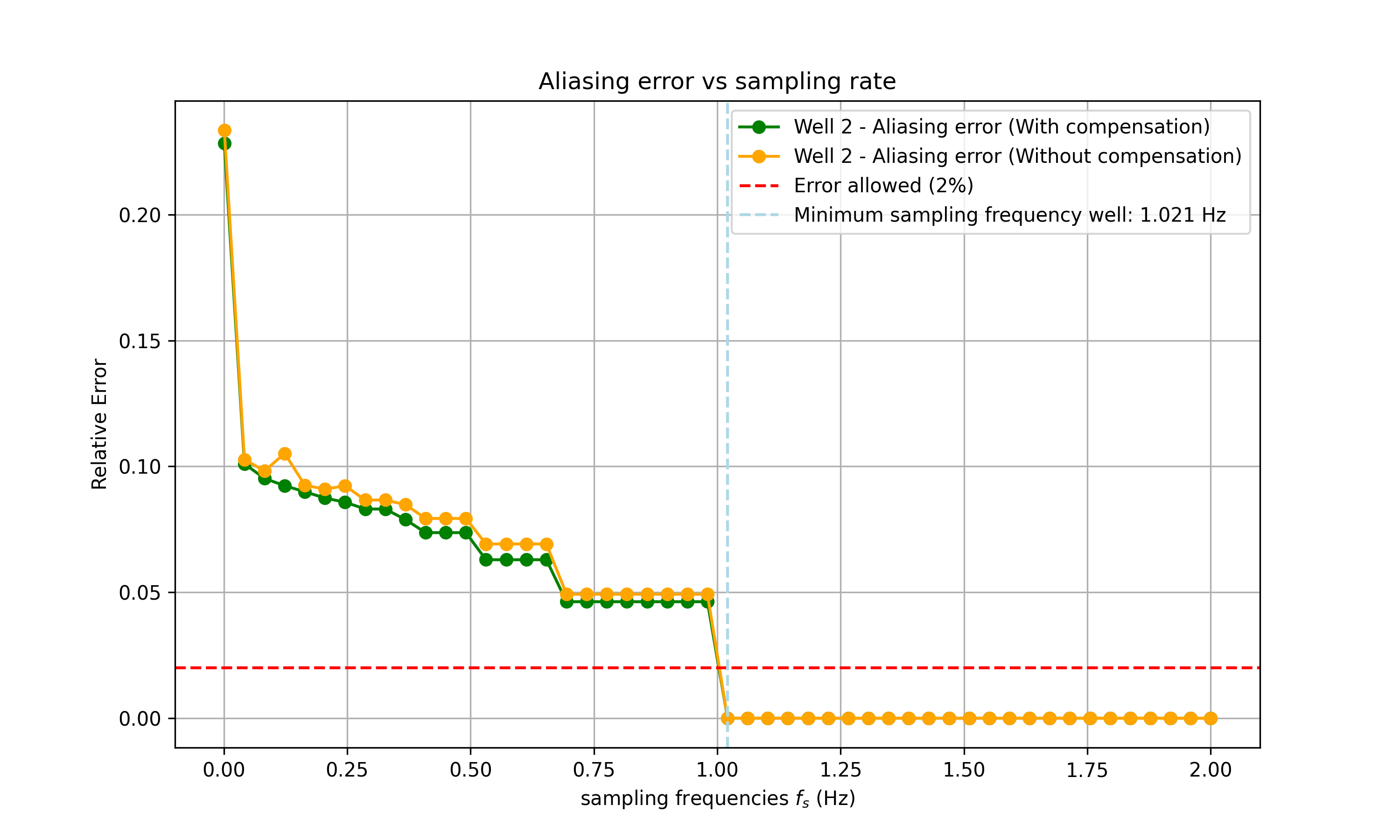}
    \caption{Well 2, relationship between L2 error and sampling rate (\(f_s\)) with compensation techniques (\(filtering + cubic reconstruction\)), showing that the compensated signal can be used at lower sampling rates than the original one.}
    \label{fig:c_well2}
\end{figure}

\begin{figure}[h!]
    \centering
    \includegraphics[width=0.8\textwidth]{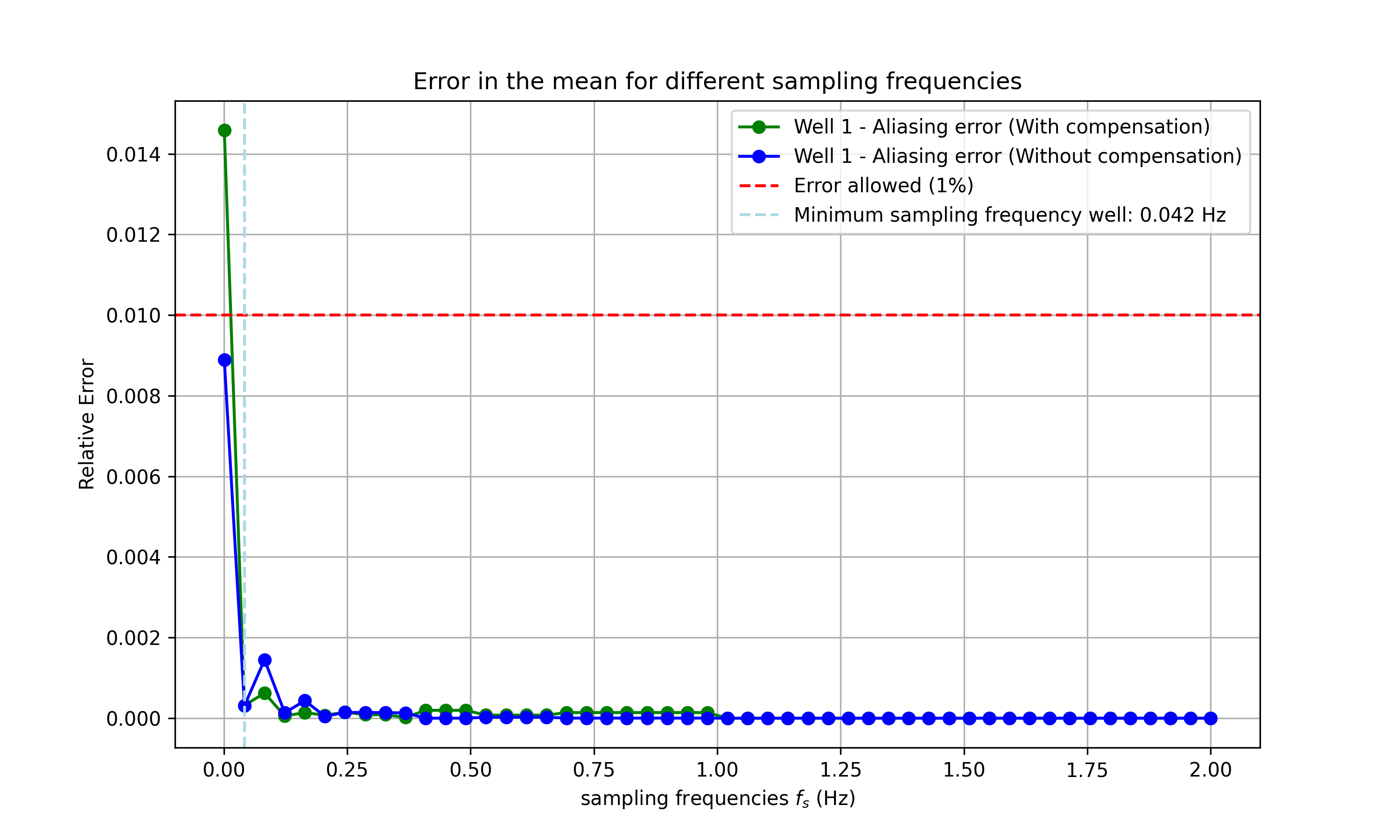}
    \caption{Well 1, relationship between the error relative to the mean and the sampling frequency (\(f_s\)) for the original signal with compensation techniques (filtering + cubic reconstruction).}
    \label{fig:c_well1_mean}
\end{figure}

\begin{figure}[h!]
    \centering
    \includegraphics[width=0.8\textwidth]{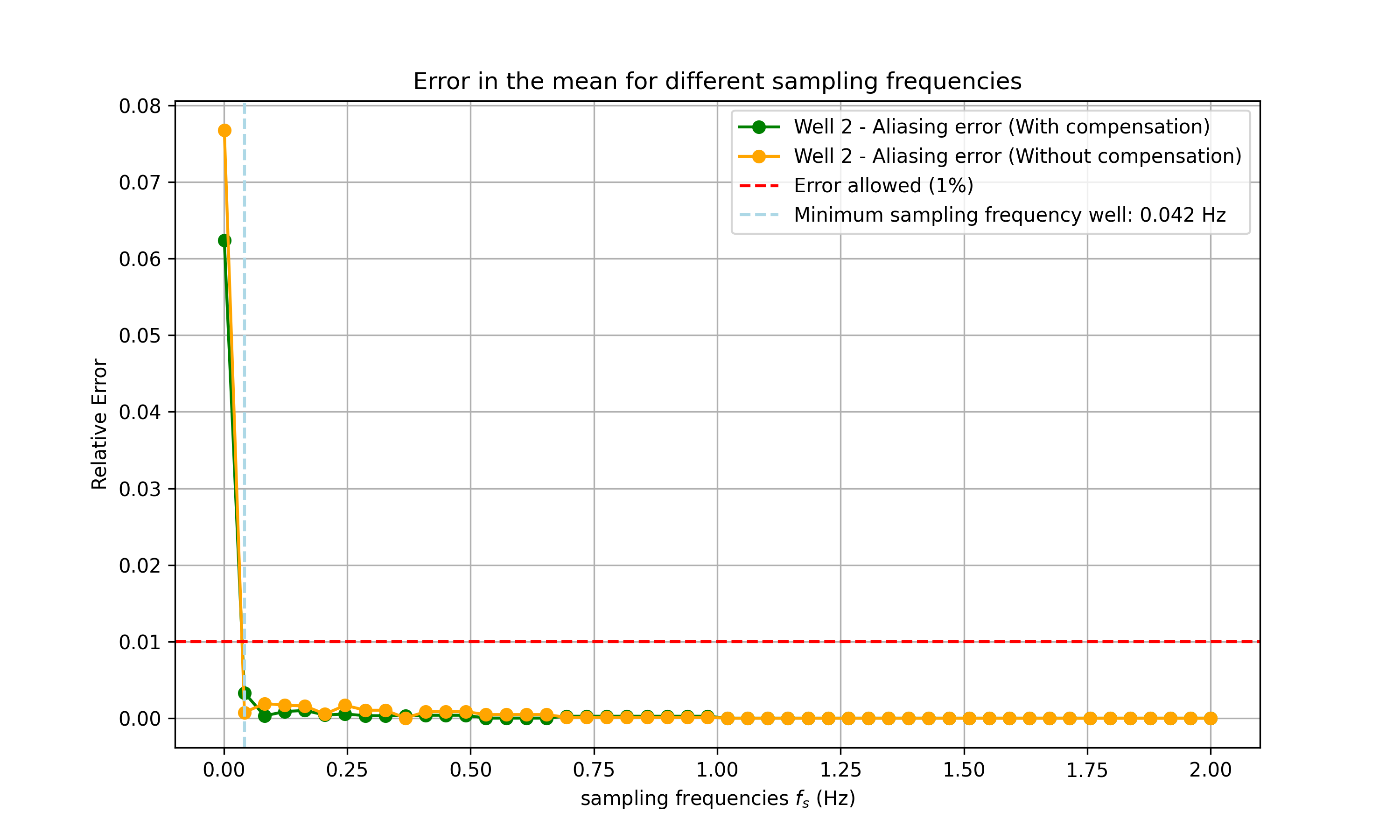}
    \caption{Well 2, relationship between the error relative to the mean and the sampling frequency (\(f_s\)) for the original signal with compensation techniques (filtering + cubic reconstruction).}
    \label{fig:c_well1_mean}
\end{figure}

\subsection{Battery Life Impact}
\subsubsection{Battery Impact Estimation}

Wireless sensor nodes rely on battery power, making energy efficiency a critical factor in industrial monitoring applications. The energy consumption per transmission is a key contributor to battery depletion, and optimizing the sampling rate directly impacts sensor longevity.

Let \( E_t \) denote the constant energy cost per transmission, independent of the sampling interval, and \( E_b \) the base consumption required to keep the sensor operational in standby mode.

The number of transmissions per hour, \( n_{\text{trans}} \), is determined by the sampling interval:

\begin{equation}
    n_{\text{trans}} = \frac{3600}{f_s}
\end{equation}

where \( f_s \) is the sampling frequency in seconds. For example, if the sensor transmits data every \(1\) second, the number of transmissions per hour is:

\begin{equation}
    n_{\text{trans,1s}} = \frac{3600}{1} = 3600
\end{equation}

Conversely, if the transmission occurs every \(5\) seconds:

\begin{equation}
    n_{\text{trans,5s}} = \frac{3600}{5} = 720
\end{equation}

The transmission ratio, comparing the high-frequency (1 Hz) to the lower-frequency (0.2 Hz) scenario, is:

\begin{equation}
    R_{\text{trans}} = \frac{n_{\text{trans,1s}}}{n_{\text{trans,5s}}} = \frac{3600}{720} = 5
\end{equation}

This means that at a 1-second interval, the sensor transmits \textbf{five times more frequently} than when using a 5-second interval, leading to proportionally higher energy consumption.

Thus, optimizing the sampling rate significantly reduces the number of transmissions and extends the sensor’s battery life, making it a crucial factor in designing energy-efficient wireless data acquisition systems.

\subsubsection{Proportional Relationship of Consumption}

The total energy consumption \(E_{\text{total}}\) in an interval can be approximated as:
\[
E_{\text{total}} = E_b + n \cdot E_t
\]
Where:
\begin{itemize}
    \item \(n\): Number of transmissions per unit of time.
    \item \(E_t\): Energy consumed per transmission.
    \item \(E_b\): Energy consumed in standby per unit of time.
\end{itemize}

With a ratio of 5 more transmissions using 1 second as an interval, the total consumption is proportionally higher. To simplify:
\[
\text{Relative consumption (5s vs 1s)} = 1 \cdot E_t \text{ (5s)} \quad \text{vs.} \quad 5 \cdot E_t \text{ (1s)}
\]
Thus, if \(E_b\) is negligible (which may be valid if transmission cost dominates), consumption in 1 second mode will be approximately 5 times higher.

\subsubsection{Battery Life Estimation}
Assuming the device battery lasts \( T \) hours with transmission at 1 second. If you change to 5 seconds, the estimated battery life will be approximately 5 times:
\[
T_{\text{5s}} = {T_{\text{1s}}} \cdot {5}
\]

For example:
If the battery lasts 2 months (1440 hours) with a transmission every 1 second changing to a transmission every 5 seconds we get:
$$T_{\text{5s}} = {1440} . {5} = {7200} \text{ hours} \approx 300 \text{days}$$
With a transmission every 5 seconds its duration is approximately 1 year, that is, we go from changing 1 battery every two months to almost annually.

\subsubsection{Practical Justification}
Based on the estimated data:
Increasing the transmission frequency from 5s to 1s would reduce battery life to approximately one-fifth.
Using a 5-second interval significantly extends battery life, which is crucial for wireless sensors in applications where replacing the battery is costly or inconvenient.
For environments where battery replacement is complex or expensive, a lower transmission frequency translates into significantly reduced operating costs and reduced sensor downtime.
However, the optimal transmission frequency should be selected based on signal quality requirements, ensuring a balance between data accuracy and power consumption.

\subsection{System Optimization}

The results obtained suggest the feasibility of developing autonomous systems that dynamically adjust sampling frequency based on well operating conditions and signal variability. Such a system could:
Monitor relative error in real-time and automatically adjust sampling frequency to optimize battery life without compromising data quality.
Implement machine learning algorithms to identify signal variability patterns and anticipate critical events.
Improve network resource management by prioritizing data based on the criticality of the information collected.
Adopting this strategy would allow for an optimal balance between data quality and energy efficiency, ensuring reliable and cost-effective monitoring in the long term.
Final Considerations and Future Directions:
This study demonstrates that it is possible to significantly reduce the sampling frequency in wireless data acquisition systems without compromising signal quality. However, careful consideration must be given to preventing aliasing effects, which can impact the ability of ML models to detect critical patterns.
The findings highlight that while stable signals allow for more aggressive sampling reduction, dynamic signals require a threshold-based approach to avoid signal degradation. A promising future direction is to develop adaptive sampling strategies that dynamically adjust the sampling frequency based on real-time signal variability, ensuring both data efficiency and signal integrity.
By striking the right balance, this methodology ensures that wireless data acquisition systems remain robust and efficient, while still providing high-quality data for advanced analytics and predictive maintenance.
In ML models applied to industrial monitoring, aliasing can introduce noise or cause the loss of valuable information, making it harder to detect early warning signs of anomalies or trends. Therefore, reducing the sampling rate should be done carefully, ensuring that the retained data continues to reflect the real patterns of the system. This study demonstrates that while an aggressive reduction in sampling frequency can save storage and transmission resources, it must be performed in a way that preserves critical features necessary for predictive analytics. The balance between data efficiency and signal fidelity is essential to maintaining the reliability of anomaly detection models.

\section{Future Work: Integrating Generative AI for Enhanced Signal Processing}
\label{sec:futurework}

This study has demonstrated the feasibility of optimizing sampling rates while preserving signal integrity for machine learning applications. However, further improvements can be achieved by integrating generative AI models to enhance data reconstruction and anomaly detection.

\subsection{Generative Models for Signal Denoising}
Generative models, such as Variational Autoencoders (VAEs) and Generative Adversarial Networks (GANs), have shown promising results in reconstructing missing or aliased signal components. These models learn the underlying distribution of signals and generate high-fidelity reconstructions, improving the accuracy of ML-based anomaly detection systems \parencite{liu_gan-based_2022, wan_adaptive_2020}.

For instance, in industrial monitoring applications, GANs have been successfully applied to reconstruct incomplete sensor readings, mitigating noise and aliasing errors \parencite{wang_gan-based_2021}. VAEs have also been explored in time-series data for predictive maintenance by learning latent representations of signal dynamics and reconstructing corrupted measurements \parencite{xu_vae_2020}.

\subsection{Adaptive Sampling with Reinforcement Learning}
Reinforcement learning (RL) can be leveraged to develop adaptive sampling strategies where an intelligent agent dynamically adjusts the sampling rate in real-time based on detected signal variability. This approach ensures optimal trade-offs between data efficiency and accuracy \parencite{panchapagesan_rl-based_2023}.

Recent studies have demonstrated that deep Q-learning and policy gradient methods can significantly reduce energy consumption in wireless sensor networks while maintaining signal fidelity \parencite{li_rl_wsn_2021}. Such approaches could be extended to industrial telemetry, allowing real-time adaptation of sampling rates to varying operational conditions.

\subsection{Multi-Modal Data Fusion}
Future research can explore how integrating additional sensor modalities—such as pressure, vibration, or acoustic data—using multimodal AI frameworks can enhance predictive maintenance capabilities in industrial settings. Recent works show that multimodal fusion with deep learning can improve fault detection in industrial processes by combining heterogeneous data sources \parencite{yang_multimodal_2022}.

A promising direction involves combining transformer-based architectures with sensor fusion to improve anomaly detection in well-testing scenarios, leveraging cross-modal correlations to refine predictions \parencite{ma_transformer_multimodal_2023}.

These advancements would allow wireless data acquisition systems to become more autonomous, efficient, and reliable, facilitating better predictive analytics in industrial environments.

\section{Conclusion}
\label{sec:conclusion}

This study has demonstrated that reducing the sampling frequency in wireless industrial data acquisition systems is feasible without significantly compromising measurement quality, provided that appropriate compensation techniques are applied. By implementing filtering and cubic interpolation, it was possible to reduce the sampling rate by up to \textbf{80\%} while maintaining a relative error below \textbf{2\%}. 

The key findings of this study include:

\begin{itemize}
    \item \textbf{Battery savings}: Lowering the sampling frequency from 1 Hz to 5-second intervals extended sensor battery life by nearly \textbf{five times}, significantly reducing operational costs in remote or hazardous industrial environments.
    \item \textbf{Storage and transmission efficiency}: An 80\% reduction in data generation leads to a substantial decrease in storage and transmission overhead, optimizing network bandwidth usage while maintaining reliable data acquisition.
    \item \textbf{Preserving critical signal features}: Without compensation techniques, aggressive downsampling introduces aliasing errors, distorting key signal patterns. However, with proper filtering and interpolation, the downsampled signal retained its essential characteristics, making it suitable for predictive analytics and anomaly detection.
    \item \textbf{Impact on machine learning models}: Many industrial monitoring applications rely on ML models to detect early warning signs of failures. If sampling rates are reduced without compensation, valuable trend information can be lost, degrading model performance. This study confirms that aliasing-free downsampling is essential for maintaining predictive accuracy.
\end{itemize}

The results highlight that while reducing sampling frequency yields clear benefits in terms of energy efficiency and data management, it must be done carefully. \textbf{A naive approach that simply reduces the sampling rate without compensation can lead to significant data loss, impairing anomaly detection and process optimization.}

Future work will focus on extending these findings by integrating \textbf{adaptive sampling strategies using machine learning}, where real-time signal variability determines optimal sampling rates dynamically. Additionally, the incorporation of **generative AI techniques** could further enhance data reconstruction, improving trend preservation and anomaly detection in industrial systems.

Ultimately, this research provides a robust framework for optimizing industrial telemetry, balancing **data efficiency, sensor longevity, and analytical reliability** to ensure high-quality decision-making in industrial operations.

\section{References}
\printbibliography

\end{document}